\documentclass[12pt]{iopart}
\usepackage[UKenglish]{babel}
\usepackage{graphicx}
\usepackage{tikz}
\usepackage{caption}
\usepackage{cases}
\usepackage{iopams}
\usepackage{xspace}
\usepackage{bbm}
\usepackage{subfig}
\usepackage{gensymb}

\newcommand{\text}[1]{\mathrm{#1}}
\usepackage{color,xcolor}

\newcommand{\elabel}[1]{\label{#1}}

\newcommand{\slabel}[1]{\label{#1}}
\newcommand{\flabel}[1]{\label{#1}}

\newcommand{\tlabel}[1]{\label{#1}}
\newcommand{\latin}[1]{{\it #1}}
\newcommand{\ie}{\latin{i.e.}\@\xspace}

\newcommand{\operatorname}[1]{\mathrm{#1}}

\renewcommand{\exp}[1]{\mathchoice{\mathrm{e}^{#1}}{\operatorname{exp}\left(#1\right)}{\operatorname{exp}\left(#1\right)}{\operatorname{exp}\left(#1\right)}}

\newcommand{\gpvec}[1]{\mathbf{#1}}

\newcommand{\evec}{\gpvec{e}}

\newcommand{\Pvec}{\gpvec{P}}

\newcommand{\plaind}{\mathrm{d}}

\newcommand{\spave}[1]{\overline{#1}}

\newcommand{\imag}{\mathring{\imath}}

\usepackage{dsfont}

\newcommand{\gpset}[1]{\mathds{#1}}

\newcommand{\canetset}[1]{{\mathchoice {\hbox{$\sf\textstyle #1\kern-0.4em #1$}}
                {\hbox{$\sf\textstyle #1\kern-0.4em #1$}}
                {\hbox{$\sf\scriptstyle #1\kern-0.3em #1$}}
                {\hbox{$\sf\scriptscriptstyle #1\kern-0.2em #1$}}}}

\newcommand{\Rset}{\gpset{R}}

\newcommand{\keyword}[1]{\emph{#1}}

\newcommand{\subc}{{\text{\scriptsize c}}}
\newcommand{\subu}{{\text{\scriptsize u}}}
\newcommand{\Lsubc}{{\!\text{\scriptsize c}}}
\newcommand{\Lsubu}{{\!\text{\scriptsize u}}}
\newcommand{\Mmatrix}{\underline{\underline{M}}}
\newcommand{\PTheovec}{\spave{\Pvec}^{\text{\scriptsize theo.}}}

\newcommand{\numClusters}{\kappa}

\newcommand{\SupplFref}[1]{\Fref{#1}}

\newcommand{\beginsupplement}{%
        \setcounter{table}{0}
        \renewcommand{\thetable}{S\arabic{table}}%
        \setcounter{figure}{0}
        \renewcommand{\thefigure}{S\arabic{figure}}%
     }

\begin{document}
        
\title{Linear stability analysis of morphodynamics during tissue regeneration in plants}

\author{Anne-Mieke~Reijne$^{1,2}$}
\author{Gunnar~Pruessner$^{1,2}$}
\author{Giovanni~Sena$^3$}
\ead{g.sena@imperial.ac.uk}
\address{$^1$Department of Mathematics, Imperial College London, London
        SW7 2AZ, United Kingdom}
\address{$^2$Centre for Complexity Science, Imperial College London,
        London SW7 2AZ, United Kingdom}
\address{$^3$Department of Life Sciences, Imperial College London, London
        SW7 2AZ, United Kingdom}

                \begin{abstract}
One of the key characteristics of multicellular organisms is the ability to establish and maintain shapes, or morphologies, under a variety of physical and chemical perturbations. 
A quantitative description of the underlying morphological dynamics is a critical step to fully understand the self-organising properties of multicellular systems. Although many powerful mathematical tools have been developed to analyse stochastic dynamics, rarely these are applied to experimental developmental biology.

Here, we take root tip regeneration in the plant model system \textit{Arabidopsis thaliana} as an example of robust morphogenesis in living tissue, and present a novel approach to quantify and model the relaxation of the system  to its unperturbed morphology. By generating and analysing time-lapse series of regenerating root tips captured with confocal microscopy, we are able to extract and model the dynamics of key morphological traits at cellular resolution. We present a linear stability analysis of its Markovian dynamics, with the stationary state representing the intact root in the space of morphological traits. 
We find that the resulting eigenvalues can be classified into two groups, suggesting the co-existence of two distinct temporal scales during the process of regeneration.

We discuss the possible biological implications of our specific results, and suggest future experiments to further probe the self-organising properties of living tissue. 
                \end{abstract}

\section{Introduction}
In most multi-cellular systems, the function of an organ or a tissue relies on its morphology and internal cellular organization. Not surprisingly, evolutionary processes tend to select mechanisms to at least partially restore an optimal tissue state that has been damaged, through controlled developmental processes known as wound repair or regeneration \cite{SanchezAlvarado:2000it,Dinsmore:1991tk}. 
From a more abstract point of view, we could refer to such a state as a stable equilibrium in the space of morphologies, and the process of regeneration as the relaxation of the system back to equilibrium after a small perturbation. Tissues exhibiting such stability to perturbation are sometimes referred to as morphologically robust \cite{Lander:2011bj}.

Since the ground-breaking and influential \keyword{theory of transformation} by D'Arcy Thompson \cite{Thompson:1917}, quantitative approaches have been successfully applied to study the changes occurring during shape formation, or \keyword{morphodynamics} \cite{Bourgine:2011vk}. In plants, rigid cell walls suppress virtually all cell migration, making the process of tissue organization much simpler to understand. Perhaps for this reason, plants have been used extensively as model systems to study morphodynamics \cite{ChickarmaneETAL:2010,RoederETAL:2011}. The very successful molecular genetics approach in developmental biology has generated a body of knowledge on the molecular mechanisms involved in pattern formation and generally in tissue organization. In plants, a number of detailed computational models have been proposed to capture the complexity of genetic networks controlling morphogenesis \cite{CoenKennawayWhitewoods:2017,Alvarez-BuyllaDavila-VelderrainMartinez-Garcia:2016} and to simulate the diversity of macroscopic forms \cite{PrusinkiewiczRunions:2012}. Nevertheless, although theoretical concepts such as \keyword{morphostate} and \keyword{ontogenic trajectories} have been explored from the point of view of dynamical systems \cite{Gutman:2004df}, a broad mathematical framework to link experimental data on morphodynamics with tissue self-organization is essentially missing.

We believe that a quantitative analysis of tissue regeneration can provide a significant step towards the development of such a framework.
Among plant organs, roots offer the advantage of strong geometrical symmetries and simple internal organization, for most part composed of just few concentric layers of cell types. The root tip still maintains an approximate rotational symmetry and is organised in a very stereotypical pattern, harbouring actively dividing cells and an apical stem cell niche \cite{PetrickaWinterBenfey:2012}. The root tip of the genetic model system \textit{Arabidopsis thaliana} offers the further advantage of being quite transparent, which makes it an ideal system for live microscopy \cite{BaessoRandallSena:2018}.

Here, we take advantage of the process of root tip regeneration in \textit{Arabidopsis} \cite{SenaETAL:2009} and combine morphometrics at cellular resolution with the theory of stochastic processes, to propose a novel quantitative approach to study dynamical perturbations of stable morphologies. We use time-lapse imaging with confocal fluorescence microscopy and measure morphological traits at cellular resolution. We propose a novel mathematical framework to describe the morphological dynamics, borrowing from field theory methods. Finally, we present a mathematical analysis of the robustness of the root morphology, or regeneration, at cellular resolution.

\section{Results}

\subsection{Distributions of cellular morphological traits}
We followed an established protocol \cite{SenaETAL:2009} to collect median longitudinal optical sections of uncut as well as cut (\ie with their tip fully excised) and thus regenerating \textit{Arabidopsis} root tips. Each root was imaged once a day, for at least 8 days (\Fref{Confocal}). 

\begin{figure} [h]
\includegraphics[width=1.0\textwidth]{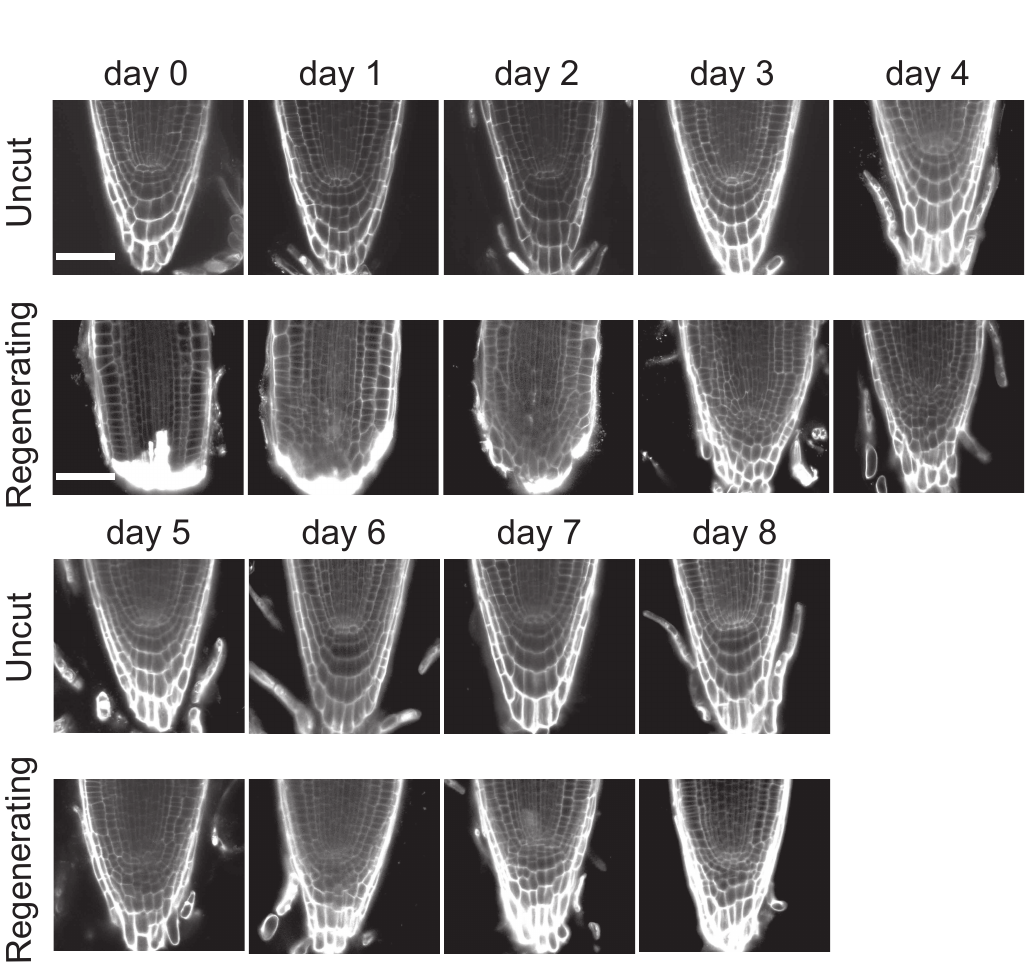}
\caption{\flabel{Confocal}
Optical longitudinal median sections of representative \textit{Arabidopsis} roots at various days after tip excision (Regenerating) or mock-treated (Uncut). Confocal microscopy with propidium iodide to counter stain cell walls. Scale bar, $50\mu m$
}
\end{figure}
We developed an original image processing routine based on the marker-controlled watershed segmentation method \cite{WatershedSegmentation:2018} to identify cells in each optical section and to measure its area, eccentricity and the orientation of its major axis. 
At each time-point, we measured these traits for all the cells in all root meristems that have been cut and imaged through regeneration, as well as in all uncut root meristems (\Fref{Box-plots}). 

\begin{figure}[h]
\centering
\includegraphics[width=1\textwidth]{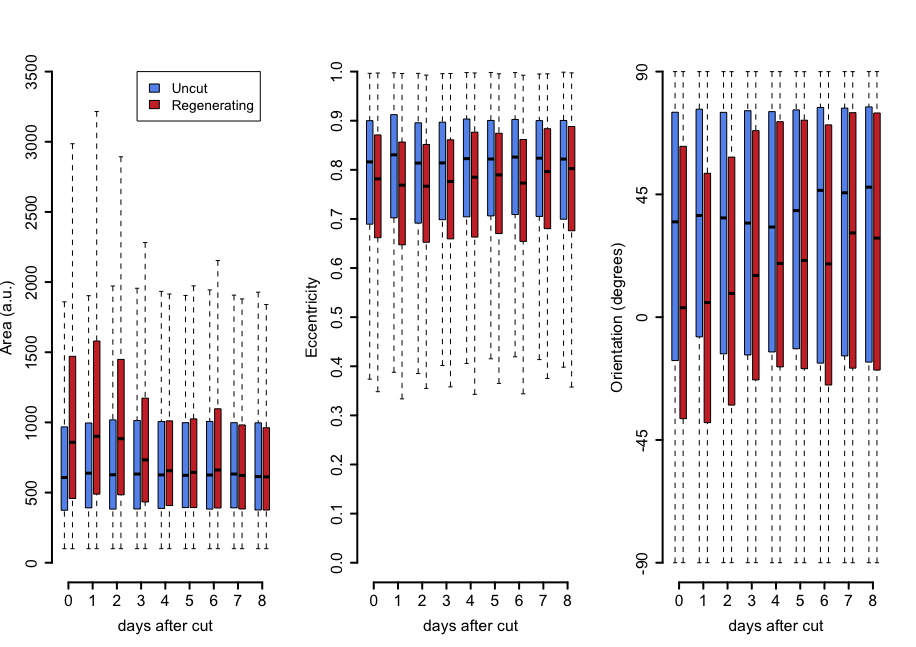}
\caption{\flabel{Box-plots}
Box-plot representation of the distribution of the measured cellular morphological traits (area, eccentricity and orientation) as a function of time, after root tip excision (Regenerating) or mock-treated (Uncut)
}
\end{figure}

\subsection{A vector field as a representation of cell identity}
In our study, cell morphology is described by three metrics, namely area, eccentricity and orientation of the major axis of the cell, encoded in the sine and cosine of the orientation angle (see \Sref{Image_processing} for detailed methods), and thus mapped in a four-dimensional space (\keyword{morphospace}). After normalising the area and the eccentricity to the interval $[0,1]$, each cell is represented by a point on the surface of a four-dimensional hyper-cylinder that has a unit circle as base.
We asked if all morphologies we identified could, in fact, be subdivided into only a few clusters, or cell types (\keyword{morphostates}). We collected area, eccentricity and orientation of all cells from a separate set of $31$uncut root tips at day $0$ ($3$ days after germination) and minimised the Davis-Bouldin index (DBi) \cite{DaviesBouldin:1979} to find the number $\numClusters $ of clusters present in the data (\SupplFref{DB_index}).

Guided by biological intuition, we chose  $\numClusters=7$ as the best solution that is both non-trivial and treatable. 
We repeated the analysis using another set of uncut roots at days $0$--$8$, and a third set composed of  
all the images we collected of uncut and regenerating roots, taken over various days after the cut. All sets identified $\numClusters=7$ as a local minimum of the DBi.
Finally, we applied the Lloyd's \cite{Lloyd:1982} and \texttt{k-means++} \cite{ArthurVassilvitskii:2007} algorithms to cluster the data points in the morphospace and to calculate the position of each cluster's centroids, which is the point to which the sum of the squared Euclidean distances
from all cells in the cluster was minimal. 
The resulting set of $\numClusters$ centroids in the four-dimensional space 
represent $\numClusters$ morphostates and thus provide a map to associate a 
morphostate with every point within that space.
The morphostate of a cell is given by the centroid closest (in terms 
of Euclidean distance) to the coordinates representing the cellular morphology.
Each cell was assigned to a morphostate 
$i\in\{1,2,\ldots,\numClusters\}$ by finding the centroid closest to the cell's position in the morphospace
(a typical map of morphostates is shown in \SupplFref{Types_map}).

In order to maintain the information about the relative position of each cell across roots, we performed \latin{ad hoc} image registration to spatially overlap all the roots (maintaining separated cut and uncut sets) at any given time-point (see \Sref{Image_processing}). For each set of registered roots at time $t$, we defined a probability $P_i (\vec{x},t)$ of finding a pixel at position $\vec{x}\in\Rset^2$ in the image plane, belonging to a cell in morphostate $i$. Thus, 
$\Pvec(\vec{x},t)=\big(P_1 (\vec{x},t),\ldots,P_{\numClusters}(\vec{x},t)\big)$ 
is a vector field representing the distribution of morphostates in a typical root  and its evolution in time is a quantitative representation of the morphological changes occurring at cellular level as the root grows (for uncut roots) or regenerates (for cut roots). An equivalent interpretation of the field $\Pvec(\vec{x},t)$ is that each cell of the typical root is a weighted superposition of morphostates, with the probability to be in a certain morphostate $i$ given by the components $P_i$ of $\Pvec$.
The spatial average of the field $\Pvec(\vec{x},t)$ is a seven-dimensional vector $\spave{\Pvec}(t)$, which we call \keyword{morphovector}, whose component $i$ represent the probability of finding anywhere a cell in morphostate $i$, at time $t$. The morphological dynamics of growing or regenerating roots is described by the evolution of its components $\spave{P}_i(t)$ in time $t$ (\Fref{Morphoevolution}). 

\begin{figure}
\centering
\includegraphics[width=1.0\textwidth]{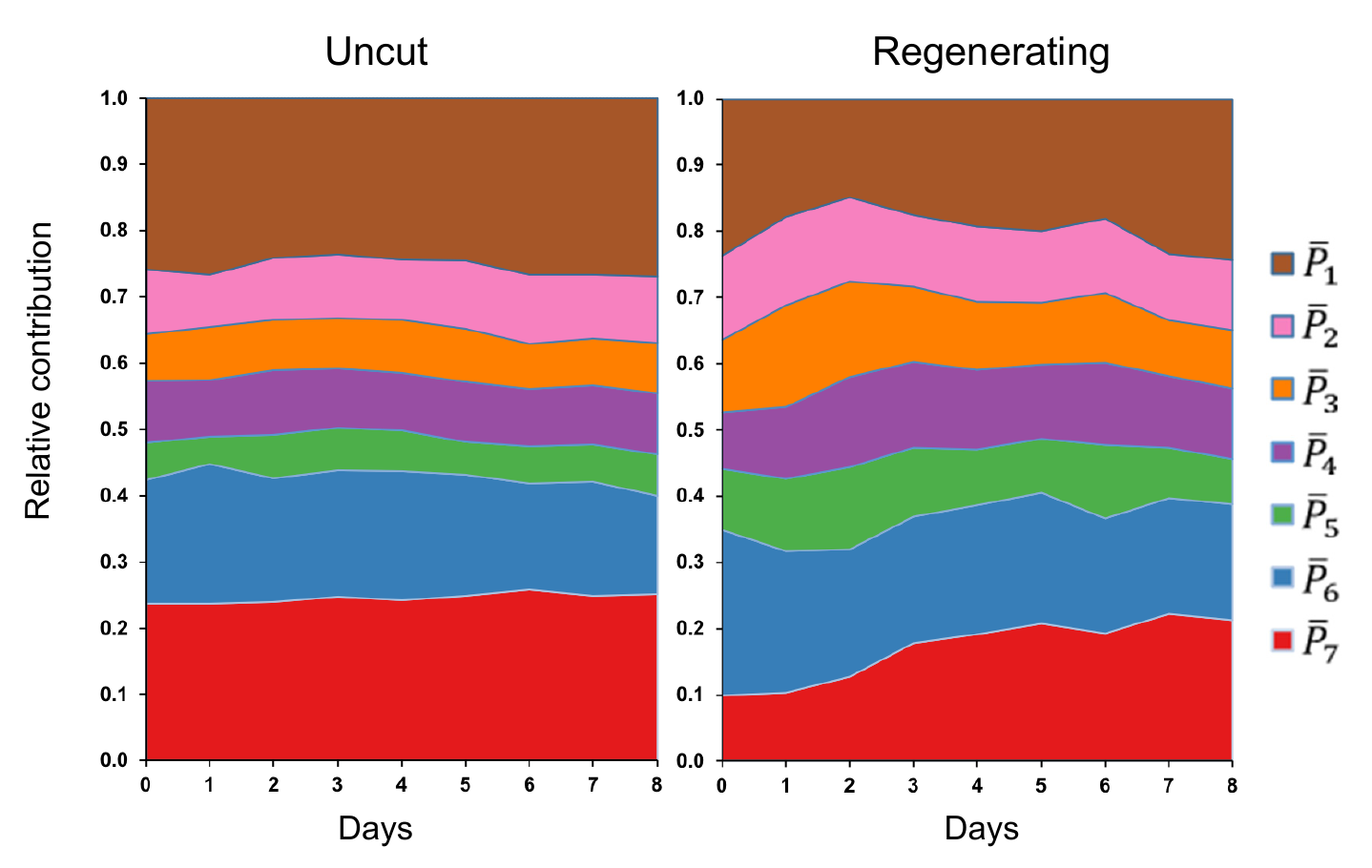}
\caption{\flabel{Morphoevolution} 
Dynamics of each of the components $\spave{P}_1(t),..,\spave{P}_7(t)$ 
of the morphovector $\spave{\Pvec}(t)$, plotted as their relative contribution and as a function of time $t$ after root tip excision (Regenerating) or mock-treated (Uncut).
}
\end{figure}

As expected, the morphovector for uncut roots shows little variation in time, while for regenerating roots the vector components show a more interesting dynamics. In essence, it appears that during regeneration the average morphological identity of the cells varies visibly as time progresses and converges back to a morphostate 
characteristic of the uncut root. Perhaps this is not surprising, as it captures the whole concept of regeneration from a morphological perspective. Less obvious is the fact that the observed variation of the morphovector is dominated by only $3$ of the $7$ morphostates 
($\spave{P}_1$, $\spave{P}_2$ and $\spave{P}_7$ in \Fref{Morphoevolution}).

\subsection{Entropy as a measurement of morphological order}
From the field $\Pvec(\vec{x},t)$ we can derive a local Shannon entropy \cite{Shannon:1960}
\begin{equation}
S(\vec{x},t)=-\,\sum^{\kappa=7}_{i=1}\,P_i(\vec{x},t)\,\ln P_i(\vec{x},t).
\end{equation}
where $S(\vec{x},t)=0$ if any cell at position $\vec{x}$ and time $t$ belongs to a single morphostate ($P_i=1$ for one, single $i$ and $P_j=0$ for all $j\ne i$) whereas $S(\vec{x},t)=
\ln \numClusters=\ln 7=1.94...$ if any cell at position $\vec{x}$ and time $t$ belongs to any of the morphostates with equal probability $1/\numClusters=1/7$, so that $\Pvec(\vec{x},t)$ is a uniform superposition of all morphostates.

The scalar field $S(\vec{x},t)$ can then be interpreted as a measure of local variability of cell morphology in a random sample of comparable roots, or as a measure of order of the cellular pattern in the tissue. Large values of $S$, \ie high entropy, indicate a morphological pattern less conserved from root to root, thus associated with \emph{disorder}. Small values of $S$, \ie low entropy, on the  other hand indicate  a more conserved tissue organization, thus associated with \emph{order}.

A visual representation of the scalar field $S(\vec{x},t)$ mapped on top of a typical root (\Fref{ShannonMaps}) shows well-defined regions of order and disorder, both in uncut and regenerating roots. Interestingly, low-order regions loosely correspond to well-characterized tissue types such as lateral root cap and epidermis ((\Fref{ShannonMaps}). Moreover, regenerating roots show a progressive re-establishment of low-entropy regions that were lost immediately after tip excision (\Fref{ShannonMaps}). 

\begin{figure}
\includegraphics[width=1.0\textwidth]{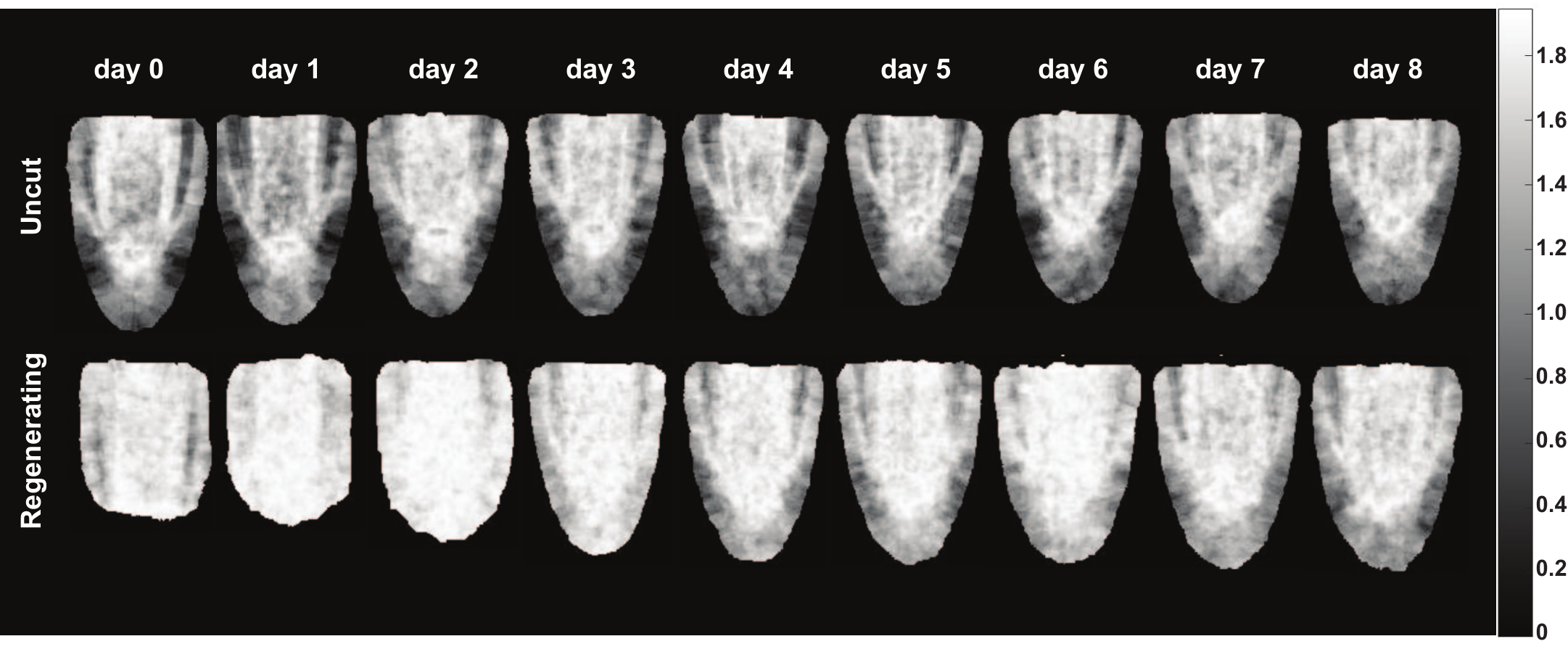}
\caption{\flabel{ShannonMaps} 
Local Shannon entropy $S(\vec{x},t)$ mapped on a typical longitudinal median section of \textit{Arabidopsis} root tip, as a function of time after tip excision (Regenerating) or mock-treated (Uncut).
}
\end{figure}

To further quantify the progress of morphological reorganization of regenerating roots, we took the spatial average $\spave{S}(t)$ of $S(\vec{x},t)$ and plotted it as a function of time (\Fref{ShannonLines}).
Its error was calculated by the Jackknife procedure \cite{Efron:1982},
\begin{equation}
\textrm{error}_{\textrm{Jackknife}}=\left[\frac{N-1}{N}\,\sum^N_{i=1}\left(\bar{S}_i-\frac{1}{N}\,\sum^N_{j=1}\bar{S}_j\right)^2\,\,\right]^{1/2},
\end{equation}
where $\bar{S}_i$ is the spatially averaged entropy when root $i$ was left out from the sample.

\begin{figure}
\centering
\includegraphics[width=0.75\textwidth]{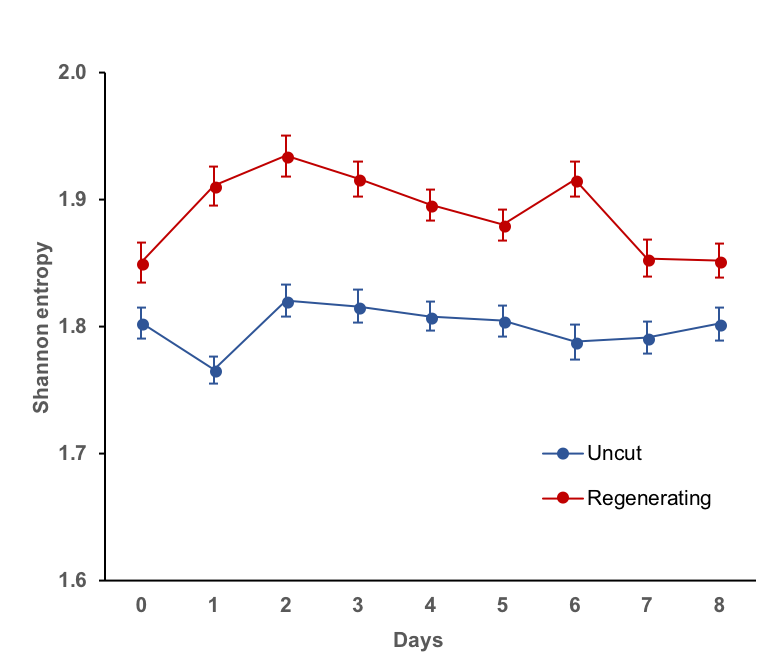}
\caption{\flabel{ShannonLines} 
Spatial average of the Shannon entropy $S(\vec{x},t)$ as a function of time after tip excision (Regenerating) or mock-treated (Uncut).
}
\end{figure}

\subsection{Linear approximation of morphological perturbation}
The biology of root tip regeneration suggests that the uncut state is a stable equilibrium in the morphological space, that the tip excision is a perturbation away from equilibrium and that the process of regeneration is the relaxation of the system back to equilibrium. The formalism and measurements described above provide us with the quantitative observables to describe mathematically the dynamics of such perturbation.

In the following, we call $\spave{\Pvec}_\Lsubu(t)$ and 
$\spave{\Pvec}_\Lsubc(t)$ the morphovectors for uncut and cut roots, respectively,\footnote{Here and in the following the subscript u and c
refer to \textbf{u}ncut and \textbf{c}ut roots respectively.} as measured on days $t\in\{0,1,\ldots,T_{\subu,\subc}-1\}$. 
Here $T_{\subu}=9$
and
$T_{\subc}=10$
denote the numbers of days uncut and cut roots have been observed for.

In a mean-field approximation, we consider the evolution of the (spatially 
averaged) morphovector $\spave{\Pvec}_\subc(t)$
as a (linear) continuous-time Markov process with master equation
\begin{equation}\elabel{dynamical_system}
\frac{\plaind}{\plaind t} \spave{\Pvec}_\Lsubc(t)
=
\Mmatrix \spave{\Pvec}_\Lsubc(t) \ ,
\end{equation}
where the matrix $\Mmatrix$ is the Markov matrix 
governing the dynamics of $\spave{\Pvec}_\Lsubc(t)$, or transition matrix.
The off-diagonal elements are thus non-negative Poissonian transition rates 
between morphostates of cells in the cut root. 
The diagonal elements, on the other hand, are determined by the constraint that columns have to sum to $0$ as a matter of probability conservation, $(1,\ldots,1)\Mmatrix=0$.
\Eref{dynamical_system} is a linearisation of a
potentially more complicated evolution of $\spave{\Pvec}$, which, here, 
we demand to obey simple linear Markovian dynamics. What follows 
therefore amounts to a linear stability analysis of the morphpodynamics.

Taylor-expanding \Eref{dynamical_system} to first order gives
\begin{equation}\elabel{Taylor_expand}
\spave{\Pvec}_\Lsubc(t+1)-\spave{\Pvec}_\Lsubc(t) = \Mmatrix \spave{\Pvec}_\Lsubc(t)
\end{equation}
and thus seven (namely number of morphostates and thus components of any morphovector $\spave{\Pvec}$) equations for $t=0,1,\ldots,T_{\subc}-2$,
resulting in $63$ equations in total, used to determine a $7\times7$ matrix.
As the diagonal elements are determined by the columns of $\Mmatrix$ summing to $0$
"only" $42$ entries are to be fitted by a regular least square fit.

The resulting matrix $\Mmatrix$ can be used 
to predict the cell frequencies $\spave{\Pvec}_\Lsubc(t+\plaind t)$ at future time
$t+\plaind t$ from those at given time $t$.
Distinguishing the theoretical expectation from the experimental measurements,
we write
\begin{equation}
\PTheovec_\subc(t+\plaind t) = \PTheovec_\subc(t) +  \Mmatrix  
\PTheovec_\subc(t) 
\plaind t
\ ,
\end{equation}
where $\PTheovec_\subc(t+\plaind t)$ is the theoretical prediction of the morphovector $\spave{\Pvec}_\Lsubc(t+\plaind t)$ at time $t+\plaind t$ and $\Mmatrix$ is the matrix obtained above. To start the integration
we set $\PTheovec_\subc(0)=\spave{\Pvec}_\Lsubc(0)$ (see also \Eref{initial} and \eref{PTheovec_integral}).

To validate the transition matrix $\Mmatrix$, we compared the components of the predicted morphovector $\PTheovec_\subc(t) $ with the experimental ones as a function of time (\Fref{Validation}). 

\begin{figure}
\centering
\includegraphics[width=0.75\textwidth]{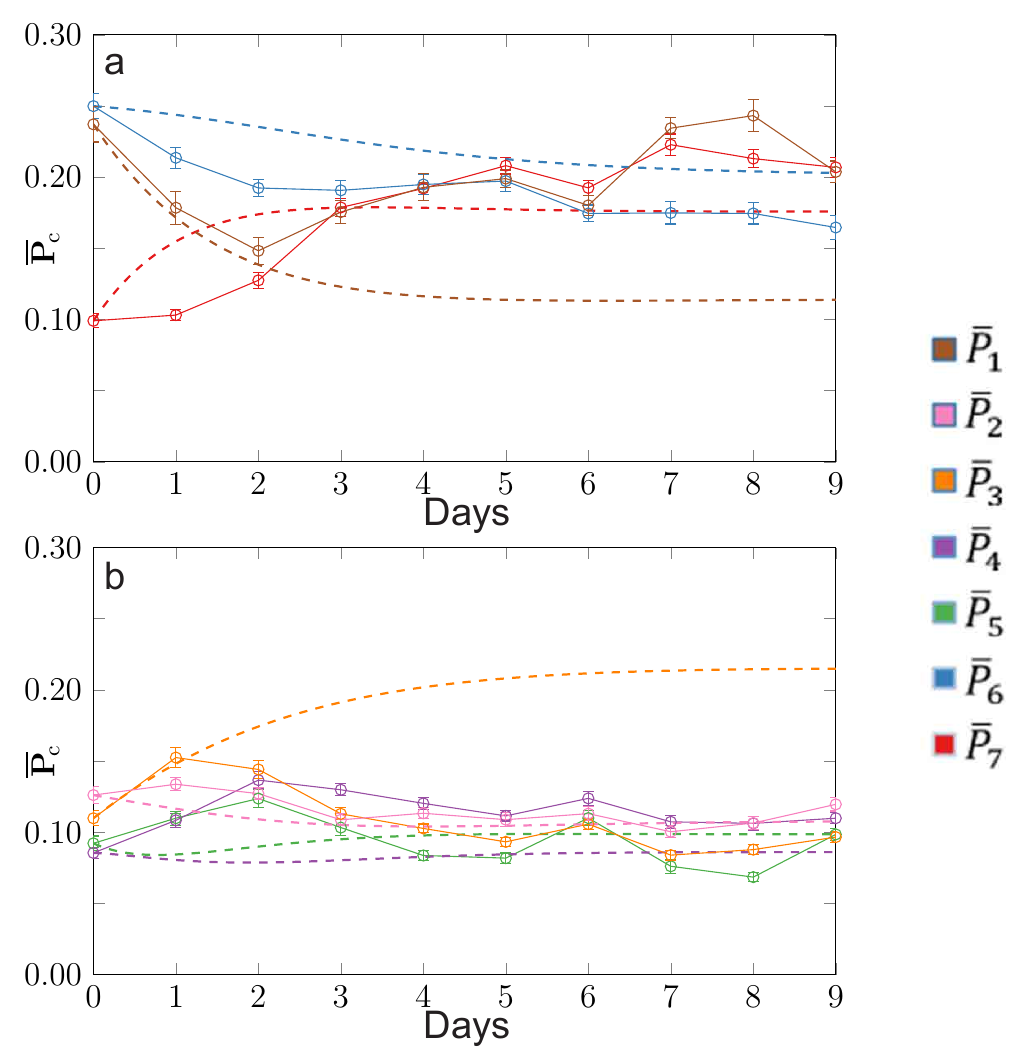}
\caption{\flabel{Validation} 
Comparison of the theoretical prediction
$\PTheovec_\subc(t)$, \Eref{PTheovec_integral}, shown as
dashed lines, to the experimental data 
$\spave{\Pvec}_\subc(t)$ shown as full lines.
Panel (a) shows the three dominant components of the vector $\spave{\Pvec}_1$, $\spave{\Pvec}_2$ and $\spave{\Pvec}_7$, with the other components shown in panel (b).
The colour code is the same as in \Fref{Morphoevolution}.
}
\end{figure}

Remarkably, two of the three most dominant  components of the morphovector $\PTheovec_\subc(t)$ appear to capture the overall features of the evolution of their experimental counterparts. 
This is quite astonishing, given the fact that the theoretical predictions are constructed by a rather brutal mean field approximation. 
However, the long-term behaviour is generally much less well captured than
the more prominent, early evolution.

\subsection{Stability analysis of the uncut morphology}
Given the transition matrix $\Mmatrix$ describing linear perturbations around the uncut morphostate, we can now proceed with a standard 
analysis \cite{vanKampen:1992} and calculate its right eigenvectors $\evec_i$ and eigenvalues $\lambda_i$,
\begin{equation}
\Mmatrix \evec_i = \lambda_i \evec_i\ .
\end{equation}
By construction, $\Mmatrix$ has one left eigenvector 
$(1,\ldots,1)$ with vanishing eigenvalue $\lambda_0=0$,
as $(1,\ldots,1)\Mmatrix=0$. 
The corresponding right eigenvector
$\evec_0$ with eigenvalue $\lambda_0$ 
represents the stationary distribution of the Markov process 
in \Eref{dynamical_system}, \ie it is the state the Markov matrix
predicts the cut root relaxes back to.  

Comparing it to the temporal average of the morphostate of the uncut root,
\begin{equation}
\spave{\Pvec}_\Lsubu^* = \frac{1}{T_\subu}\sum_{t=0}^{T_\subu-1} \spave{\Pvec}_{\Lsubu}(t)
\end{equation}
we found a good qualitative correspondence between the weight of
the seven different morphostates. This suggests that the period of
$T_c$ days, that fed into the analysis, suffices to produce a 
rough estimate of the asymptotic distribution of morphostates.

In general, all other eigenvectors $\evec_i$ and eigenvalues $\lambda_i$ of $\Mmatrix$
are complex. As $\Mmatrix$ is real, their complex conjugates are linearly independent eigenvectors themselves.
If $\Mmatrix$ has 
eigenvalues $\lambda_i$ and
eigenvectors $\evec_i$ with $i=1,2,\ldots,6$ that span the entire subspace of $\Rset^7$ 
of normalised morphostate vectors, namely those orthogonal to the left eigenvector
$(1,\ldots,1)$,
\begin{equation}\elabel{initial}
\spave{\Pvec}_\Lsubc(0)=\sum_{i=1}^6 q_i \evec_i \ ,
\end{equation}
\Eref{dynamical_system} can be integrated to give 
\begin{equation}\elabel{PTheovec_integral}
\PTheovec_\subc(t)=\sum_{i=1}^6 \exp{\lambda_i t} q_i \evec_i
\end{equation}
with characteristic relaxational time scales $-1/\Re(\lambda_i)$ derived from the real part $\Re(\lambda_i)$ of the eigenvalues $\lambda_i$ (\Tref{timescales}). 
They represent the time it takes for a perturbation to decay by a factor $1/e$.
The imaginary part of the eigenvalues, on the other hand, give rise to a superimposed oscillatory behaviour with a period of
$2\pi/\Im(\lambda_i)$.
The eigenvalues come in three pairs of complex conjugates and any linear combination \eref{initial} that is real at time $t=0$ will remain real under the evolution \eref{PTheovec_integral}.

\begin{table}
\caption{\tlabel{timescales}Eigenvalues of $\Mmatrix$ and their negative inverse real part, which provide the time scales of the regeneration in units of days.}
\begin{indented}
\item[]\begin{tabular}{@{}llll}
\br
Eigenvalue 
& numerical value 
& $-1/\Re(\lambda_i)$ in days
& $2\pi/|\Im(\lambda_i)|$ in days\\
\mr
$\lambda_1$ & $-0.931\ldots + 0.122\ldots \imag\quad$ & $1.073\ldots$ & $51.40\ldots$ \\
$\lambda_2$ & $-0.931\ldots - 0.122\ldots \imag\quad$ & $1.073\ldots$ & $51.40\ldots$ \\
$\lambda_3$ & $-0.689\ldots + 0.349\ldots \imag\quad$ & $1.451\ldots$ & $17.97\ldots$ \\
$\lambda_4$ & $-0.689\ldots - 0.349\ldots \imag\quad$ & $1.451\ldots$ & $17.97\ldots$ \\
$\lambda_5$ & $-0.240\ldots + 0.307\ldots \imag\quad$ & $4.162\ldots$ & $20.46\ldots$ \\
$\lambda_6$ & $-0.240\ldots - 0.307\ldots \imag\quad$ & $4.162\ldots$ & $20.46\ldots$ \\
\br
\end{tabular}
\end{indented}
\end{table}

We find that all eigenvalues have negative real parts, indicating that every initial state $\spave{\Pvec}_\Lsubc(0)$ will eventually relax to the stationary state. 
This is a property of the linear Markov dynamics imposed in \Eref{dynamical_system}.
What is less trivial is that the relaxation time-scales derived from $\Mmatrix$ are $1.07\ldots$ days, $1.45\ldots$ days and $4.16\ldots$ days, clearly indicating two types of relaxations: one rapid with characteristic temporal scale around one day, and a slower one with a scale of around four days. 
The oscillatory periods, on the other hand, are at least $17$ days long. They all stretch well beyond the observation period, so that, given the exponential decay of the relative amplitude, it is not realistic to expect to observe the oscillation experimentally.

\section{Discussion}
We have taken an unconventional approach in analysing quantitative morphometric data extracted from regenerating plant tissue. We collected morphometric data at cellular resolution during the regeneration of cut root tips back to their uncut morphology, and proposed an innovative way to  describe the morphological \textit{state}  of each cell, as a weighted superposition of fundamental states. We believe this is a representation that is quite effective and useful from a biological perspective, as it highlights the known fact that developmental and metabolic cellular states are fundamentally noisy. So it makes more sense to have a mathematical framework that associates each position in the tissue with a set of basic cell-types, each weighted with an evolving probability of actually occurring, instead of an over-simplified view where each position is defined at any single time by a unique cell type. From this point of view, our proposed \textit{morphovector} is designed to capture the intrinsic morphological variability observed at cellular level in any living tissue.

Moreover, we could fit a theoretical transition matrix to describe the dynamics of such a morphovector, 
and showed that it can indeed be described as a Markovian relaxation of a morphological perturbation.

From the point of view of  developmental biology, this can be interpreted as a robust morphology (stable fixed point) that resists physical damages (perturbations) by regenerating (relaxing back to) the original shape, regardless of the way it is cut, or in which direction it is pushed within the space of morphologies. 

We can go further than that. Our quantitative analysis of the morphological robustness offers information about the characteristic dynamics of the tissue regeneration. For example, the real part of the eigenvalues gives us a temporal scale of the relaxation, with respect to the direction of the corresponding eigenvector. 
Each eigenvector can be interpreted as a special direction in the morphospace, so that when the tissue is perturbed along that direction, we can predict it will regenerate with a characteristic temporal scale given by the corresponding eigenvalue. 
Interestingly, our results indicate two classes of temporal scales in the specific case of \textit{Arabidopsis} root tip regeneration: one  of about one day and another of about four days. 
That is a surprising and intriguing result, because it suggests the co-existence of two potentially distinct developmental processes, each characterised by a quite different temporal scale. At the moment we could only speculate what these processes might be, but in principle it should be possible to isolate genetic mutations where the two temporal scales are uncoupled. So, for example, one class of mutants could exhibit a regeneration dynamics characterized only by the 1-day scale, while another class would be characterized only by the slower process over a 4-day temporal scale. Crucially, we might also predict that a wild-type root could in principle be perturbed (by cuttings or other means) exactly along the eigenvector corresponding to a regeneration with a single 1-day or 4-day temporal scale. Finally, the imaginary components of the eigenvalue might suggest superimposed oscillatory trajectories during relaxation in the morphospace. Although in the system studied the oscillations would be on a very different time scale and are in effect not observable, we should not rule out the possibility that in other morphodynamic systems (organisms) these might become visible.

It should be noted that in this study we used only morphological data from root tips that actually regenerated and that we imposed Markovian dynamics. This means that we forced ourselves to explicitly study the morphodynamics in the basin of attraction of the single, stable fixed point 
representing the regenerated root tip. In fact, the probability of successful regeneration for this kind of physical damage in \textit{Arabidopsis} roots is a function of the position of the incision and is always less than unity \cite{SenaETAL:2009}. Analogous morphometric data could be collected for those roots that did not successfully regenerate.
In morphospace this may correspond to a complete loss of a well-defined position if no meaningful morphometric data can be extracted from the roots or it may correspond to another fixed point. To model this, we would need
to allow for a more complicated dynamics beyond simple (linear) Markovian evolution.

In the presence of another fixed point, what kind of perturbation would push the system out of the basin of attraction of the first fixed point and towards the second one? 
And what would be the characteristic temporal scales of the morphodynamics towards the second fixed point?
At the moment, these remain open questions that we plan to address experimentally in future works. 

To our knowledge, this is the first time that such a technique is applied to morphodynamics in the context of developmental biology, and we suggest it can be developed as a standard method to predict robustness of morphologies. We suggest that this approach could be applied to any kind of regenerating organism, where time-lapse images of morphological dynamics are available. In more general scenarios, where the eigenvalues take both negative and positive values, this method could be used to predict which perturbations (along which eigenvector in the morphospace) are more likely to produce a rapid regeneration (relaxation with large negative eigenvalues), slow regeneration (eigenvalues with small negative values) or no regeneration at all (positive eigenvalues). 

\section{Methods}
\slabel{Methods}
\subsection{Plant material}
The plants used in this study were all  wild-type \textit{Arabidopsis thaliana} individuals of Columbia (Col-0) ecotype. Seed sterilization and synchronization were performed following standard procedure \cite{SenaETAL:2009}. First, seeds were imbibed in water and stored at 4\degree C for 2 days. The seeds were then sterilized in 50\% household bleach (Sodium Hypocholrite, 5\%) and 0.0005\% Triton X-100 (Sigma, T8787) for 3 minutes and rinsed six times in sterile water. Under sterile conditions the seeds were transferred to a standard 0.8\% agar solid MS medium 
(4.4 g/l, or 1$\times$, MS basal salts [Sigma, M5519], 
 0.5\% sucrose, 
 0.05\% MES hydrate [Sigma, M8250], 
 adjusted to pH 5.7 with 1M KOH before adding 
 0.8\% agar [Sigma, A5040]) 
 and germinated vertically in a growth chamber 
 (23\degree C, 120 $\mu\textrm{mol}\,\textrm{m}^{-2}\textrm{s}^{-1}$ 
 light intensity on a 16h/8h light/dark cycle).

\subsection{Root excision}
The root tip excision was performed on primary roots 3 days post-germination, following an established protocol \cite{SenaETAL:2009}. Plants were transferred to 5.0\% agar solid medium 
(4.4 g/l, or 1$\times$, MS basal salts [Sigma, M5519], 
0.5\% sucrose, 
0.05\% MES hydrate [Sigma, M8250], 
adjusted to pH 5.7 with 1M KOH before adding
5.0\% agar [Sigma, A5040]) 
and the root tips were cut by hand with sterile 27G needles (Sterican) under a dissecting stereo-microscope (Nikon SMZ1000 at 180$\times$ magnification). The plants were then returned to a 0.8\% agar solid medium and placed in the growth chamber. The excisions were performed at 120 $\mu$m from the tip, proximal to the quiescent centre (QC), with an estimated error of $\pm \textrm{20}\mu \textrm{m}$. 

\subsection{Microscopy}
Individual plants were imaged once a day for at least 9 days, 
following an established protocol \cite{SenaETAL:2009}. Before each imaging session, each root was stained with filter-sterilized 10 $\mu$g/ml propidium iodide (Sigma, P4170) for 30 seconds to 3 minutes (shorter time for days closer to the root tip excision), briefly washed in sterile water and mounted in sterile water on sterile microscope slides. Stained root tips were imaged using a Leica SP5 laser scanning confocal microscope with a 63X water immersion objective, with excitation at 488 nm and emission between 579 nm and 698 nm. After imaging, the roots were transferred from the microscope slide to the 0.8\% agar solid medium and returned to the growth chamber.
       
\subsection{Image processing}\label{Image_processing}
All image processing routines were performed in MATLAB \cite{MATLAB}.

\subsubsection*{Segmentation.}
\slabel{segmentation_methods}
To identify single cells, each root was \textit{segmented} in three steps: removing the background, minimizing noise and applying marker-controlled watershed segmentation (\SupplFref{Segmentation}). 
The background was removed by hand, by drawing a mask on top of the root and setting all pixels outside of the mask to zero intensity. The noise was minimized by dividing the image by a blurred image of itself, obtained applying a filter with a kernel sized 101$\times$101 pixels. The marker-controlled watershed segmentation is composed of five steps \cite{WatershedSegmentation:2018}: edge recognition through the Sobel method, identification of foreground markers (intensity maxima inside each cell) through opening-closing by reconstruction, identification of background markers (cell walls) by iteratively applying a global threshold, eliminating identified cells and increasing the threshold by a small step. In addition, the foreground marker list was updated with an eroded version of the cell walls found in the previous step. Finally, the foreground and the background markers were added to the image as regional minima, and a watershed-based segmentation was performed. The segmented objects were identified as cells only if the surface area of the cells was between 100 and 10000 pixels.  

\subsubsection*{Morphological traits.}
\slabel{morpho_methods}
For each identified cell, three morphological traits were measured: area, eccentricity and orientation. 
The cell \emph{area} was defined by the number of pixels this cell encapsulated. To derive a coordinate from it, the minimum cell size observed was mapped to $0$ and the maximum to $1$.
The cell \emph{eccentricity} was defined as the eccentricity of an ellipse with the same second-moments as the cell, and then normalised between $0$ (minimum measured eccentricity) and $1$ (maximum measured eccentricity).

The cell \emph{orientation} was defined as the angle $\theta\in[0,\pi)$ between the middle line of the root and the major axis of the ellipse with the same second-moments as the cell (\SupplFref{orientation}(a)).

The cell orientation ought to be continuous and periodic with a period of $\pi$
(\Fref{orientation}(b)). 
The angle $\theta$ returned by the code as the orientation 
of a cell lies on the interval $[0,\pi)$. Two seemingly very different 
orientations of $\theta=\epsilon$ and $\theta'=\pi-\epsilon$ are in fact 
identical in the limit $\epsilon\to0$.
For the clustering to consider these two orientations to be similar for small $\epsilon$, a suitably periodic, continuous mapping of the angle is needed. In the present work, we have used 
the two coordinates 
$c=\cos(2\theta)$ and $s=\sin(2\theta)$. 
These two coordinates are located on a unit circle.

To account for the presumed symmetry of the root about its central axis, the orientation was recorded in conjunction with the relative position of the cell's centroid in relation to the central axis. As far as 
the orientation enters into the determination of a cell's morphostate, a cell located to the
right of the central axis and having angle $\theta$ with it is 
considered to have the same orientation as 
a cell to the left with angle $\theta'=\pi-\theta$ (\SupplFref{orientation}(c)).
On the other hand, 
cell orientations of two cells on the same side with angles $\theta$ and $\theta'=\pi-\theta$ respectively should generally be considered as different.
The mirror symmetry is thus accounted for by recording the original angle $\theta$ for cells to the right, but $-\theta'$ for cells on the left. If $\theta'=\pi-\theta$ (\SupplFref{orientation}(c)) the angles recorded are therefore $\theta$ and $\theta-\pi$, which means that both are effectively identical, given the periodicity $\pi$ of the orientation.

Together, the area and the eccentricity  both on unit intervals $[0,1]$, and the coordinates $(c,s)$ on
a unit circle, place the set of morphological traits of a cell on
the surface of a hyper-cylinder with a unit circle as the base.

\subsubsection*{Registration.}
\slabel{registration_methods}
To create a coordinate reference, a middle line was drawn on each root by linking the very root tip to the quiescent centre (QC) and continuing proximally in the middle of the root and parallel to its main axis (\SupplFref{Registration}(a)). The position of a cell was then defined with a  radial coordinate $r$ given by the distance of the centroid of the cell to the root middle line just defined, and with a height coordinate $h$ given by the distance between the QC and the centroid's projection on the middle line (positive when proximal and negative when distal). Radial coordinates $r$ were then normalized to $\overline{r}=\frac{r}{R}$, with $R$ the maximum radius of the root.
Each root was rotated and translated such that the middle line was vertical (purple line in \SupplFref{Registration}(a)) and that all QCs would overlap in the same position. The entire middle line (purple and green in \SupplFref{Registration}(b)) was then drawn vertical and each centroid was translated accordingly, to maintain their original $(\overline{r},h)$ coordinates.  Finally, new virtual cells were created using Voronoi tessellation \cite{Voronoi:1907} around the centroids (\SupplFref{Registration}(c)). This registration step was important for the calculation of the morphovector $\Pvec(\vec{x},t)$ described above, 
as each Voronoi cell in the registered image is associated with a centroid
of a cell in the original image, so that each \emph{pixel} $\vec{x}$ 
within a Voronoi cell can be associated with the morphological traits
of that cell. These morphological traits are not distorted by the registration
as they have been taken prior to it.

\section{References}
\bibliography{main}

\providecommand{\newblock}{}
\begin{thebibliography}{10}
\expandafter\ifx\csname url\endcsname\relax
  \def\url#1{{\tt #1}}\fi
\expandafter\ifx\csname urlprefix\endcsname\relax\def\urlprefix{URL }\fi
\providecommand{\eprint}[2][]{\url{#2}}

\bibitem{SanchezAlvarado:2000it}
Sanchez~Alvarado A 2000 {\em BioEssays\/} {\bf 22} 578--590

\bibitem{Dinsmore:1991tk}
Dinsmore C~E and of~Zoologists A~S 1991 {\em {A History of regeneration
  research : milestones in the evolution of a science}\/} (Cambridge University
  Press)

\bibitem{Lander:2011bj}
Lander A~D 2011 {\em Cell\/} {\bf 144} 955--969

\bibitem{Thompson:1917}
Thompson D~W 1917 {\em {O}n growth and form\/} (Cambridge, UK: Cambridge
  University Press)

\bibitem{Bourgine:2011vk}
Bourgine P and Lesne A 2011 {\em Morphogenesis\/} (Berlin, Germany:
  Springer-Verlag)

\bibitem{ChickarmaneETAL:2010}
Chickarmane V, Roeder A~H~K, Tarr P~T, Cunha A, Tobin C and Meyerowitz E~M 2010
  {\em Annu. Rev. Plant Biol.\/} {\bf 61} 65--87 ISSN 1543-5008

\bibitem{RoederETAL:2011}
Roeder A~H~K, Tarr P~T, Tobin C, Zhang X, Chickarmane V, Cunha A and Meyerowitz
  E~M 2011 {\em Nat. Rev. Mol. Cell Biol.\/} {\bf 12} 265--73 ISSN 1471-0072

\bibitem{CoenKennawayWhitewoods:2017}
Coen E, Kennaway R and Whitewoods C 2017 {\em Development\/}  4203--4213

\bibitem{Alvarez-BuyllaDavila-VelderrainMartinez-Garcia:2016}
Alvarez-Buylla E, D{\'{a}}vila-Velderrain J and Mart{\'{\i}}nez-Garc{\'{\i}}a J
  2016 {\em BioScience\/} {\bf 66} 371--–383

\bibitem{PrusinkiewiczRunions:2012}
Prusinkiewicz P and Runions A 2012 {\em New Phytologist.\/} {\bf 193} 549--69
  ISSN 0028-646X

\bibitem{Gutman:2004df}
Gutman D~R and Belmonte J~C~I 2004 {\em BioEssays\/} {\bf 26} 405--412

\bibitem{PetrickaWinterBenfey:2012}
Petricka J, Winter C and Benfey P 2012 {\em Annu Rev Plant Biol.\/} {\bf 63}
  563--90 ISSN 1543-5008

\bibitem{BaessoRandallSena:2018}
Baesso P, Randall R and Sena G 2018 {\em Methods Mol. Biol.\/} {\bf 1761}
  145--163

\bibitem{SenaETAL:2009}
Sena G, Wang X, Liu H, Hofhuis H and Birnbaum K 2009 {\em Nature\/} {\bf 457}
  1150--1153 ISSN 0028-0836

\bibitem{WatershedSegmentation:2018}
MathWorks 2018 Marker-controlled watershed segmentation [Online; accessed 18
  Apr 2018]
  \urlprefix\url{https://www.mathworks.com/help/images/examples/marker-controlled-watershed-segmentation.html}

\bibitem{DaviesBouldin:1979}
Davies D~L and Bouldin D~W 1979 {\em IEEE Trans. Pattern Anal. Mach. Intell.\/}
  {\bf 1} 224--227 ISSN 0162-8828

\bibitem{Lloyd:1982}
Lloyd S 1982 {\em IEEE Trans. Inf. Theory\/} {\bf 28} 129--137 ISSN 0018-9448

\bibitem{ArthurVassilvitskii:2007}
Arthur D and Vassilvitskii S 2007 k-means++: The advantages of careful seeding
  {\em Proceedings of the Eighteenth Annual ACM-SIAM Symposium on Discrete
  Algorithms\/} SODA '07 (Philadelphia, PA, USA: Society for Industrial and
  Applied Mathematics) pp 1027--1035 ISBN 978-0-898716-24-5
  \urlprefix\url{http://dl.acm.org/citation.cfm?id=1283383.1283494}

\bibitem{Shannon:1960}
Callen H~B 1960 {\em Thermodynamics and an introduction to thermostatistics\/}
  (New York, NY, USA: John Wiley \& Sons)

\bibitem{Efron:1982}
Efron B 1982 {\em The Jackknife, the Bootstrap and Other Resampling Plans\/}
  (Philadelphia, PA, USA: SIAM)

\bibitem{vanKampen:1992}
{van Kampen} N~G 1992 {\em Stochastic Processes in Physics and Chemistry\/}
  (Amsterdam, The Netherlands: Elsevier Science B. V.) third impression 2001,
  enlarged and revised

\bibitem{MATLAB}
MATLAB 2017 {\em version 9.3 (R2017b)\/} (Natick, Massachusetts: The MathWorks
  Inc.)

\bibitem{Voronoi:1907}
Voronoi G 1907 {\em J. Reine Angew. Math.\/} {\bf 133} 97--178

\end{thebibliography}

\newpage
\beginsupplement
\section*{Supplementary figures}

\begin{figure}[h]
\centering
\includegraphics[width=0.75\textwidth]{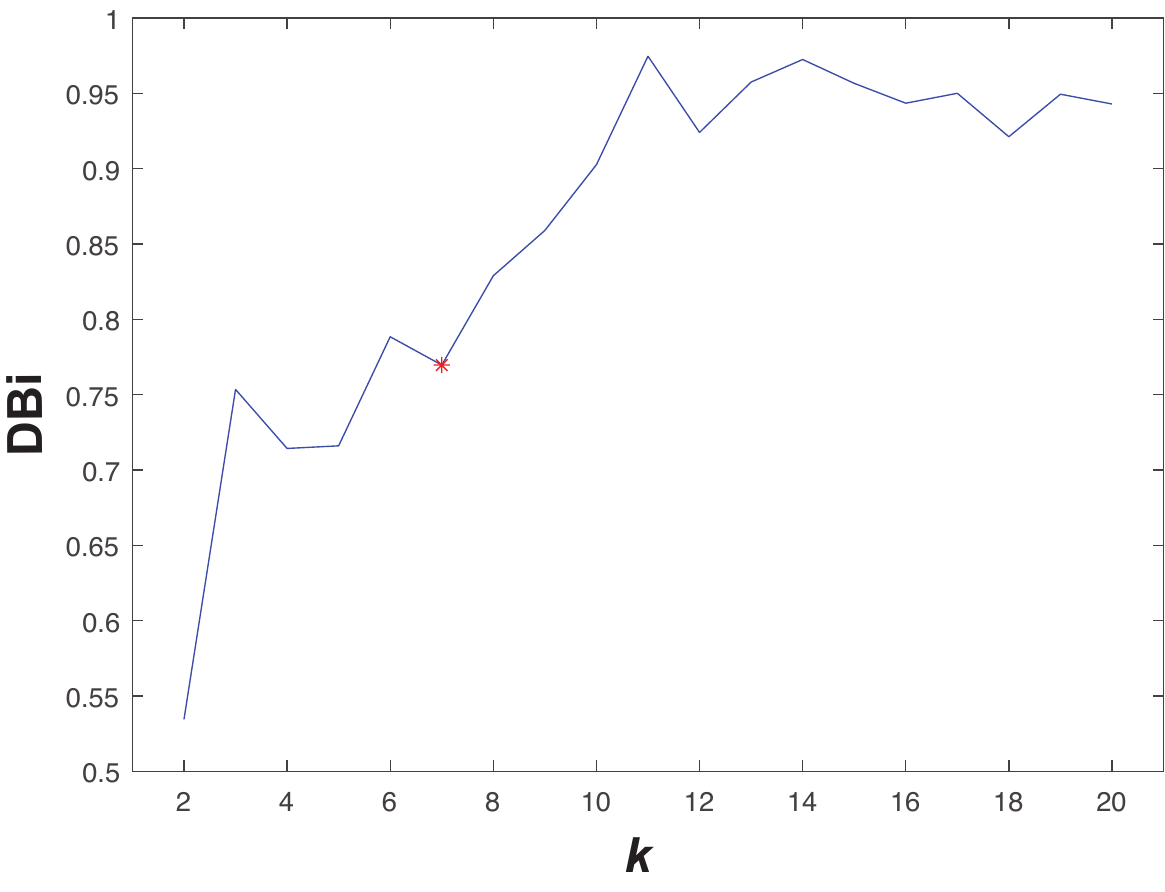}
\caption{\flabel{DB_index}
Davis-Bouldin index (DBi) calculated from our full dataset, for $k=2,...,20$. The minimum at $k=7$ was chosen as a non-trivial solution with sufficient biological significance.
}
\end{figure}

\begin{figure}[h]
\centering
\includegraphics[width=0.5\textwidth]{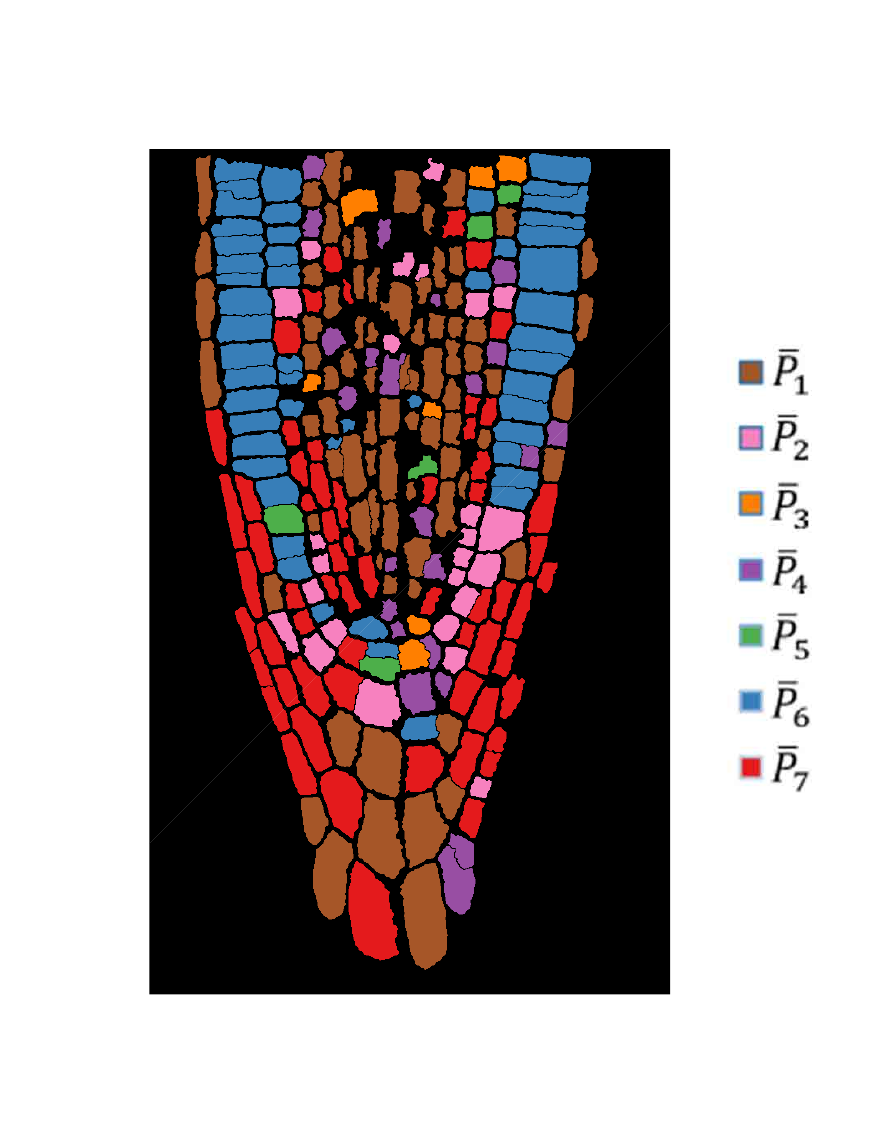}
\caption{\flabel{Types_map}
Typical root map of morphostates in a single root, colour-coded as in \Fref{Morphoevolution}
}
\end{figure}

\begin{figure}
\centering
\includegraphics[width=0.75\textwidth]{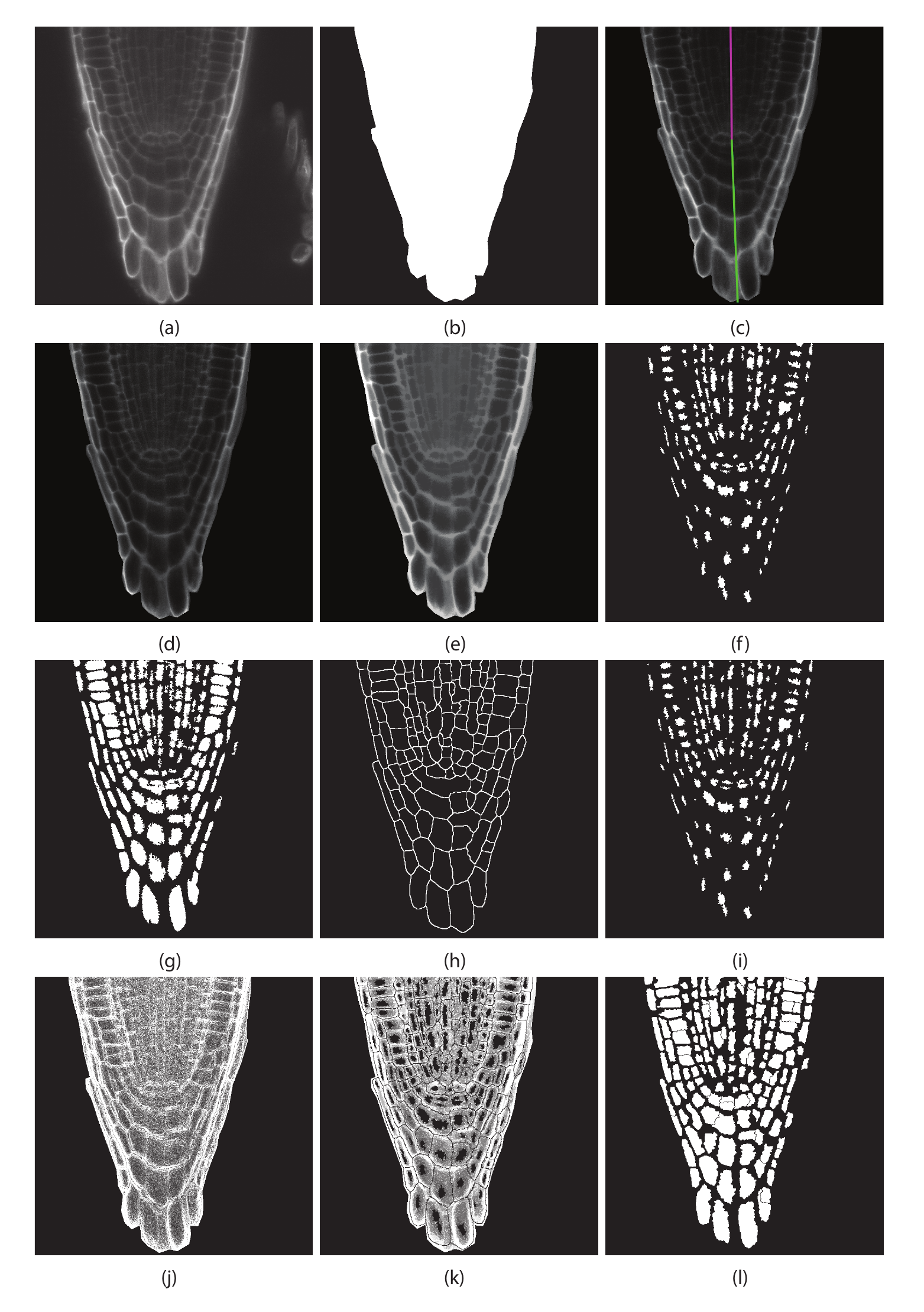}
\caption{\flabel{Segmentation}
Steps implemented in MATLAB to segment single cells from confocal images. a) Raw image. (b) Mask to identify root. (c) The root "middle line", composed of a QC-to-proximal line (purple) and a QC-to-distal one (green). (d) Image after noise minimization. (e) Image after opening-closing by reconstruction. (f) Local maxima used as foreground markers. (g) Expanded foreground markers. (h) Borders between forward markers, used as background markers. (i) Modified foreground markers with the background markers. (j) Segmentation image, defined by edge in the de-noised image. (k) Modified segmentation image with foreground and background markers as regional minima. (l) Final segmented cells.
}
\end{figure}

\begin{figure}
\centering(a)\includegraphics[height=5cm]{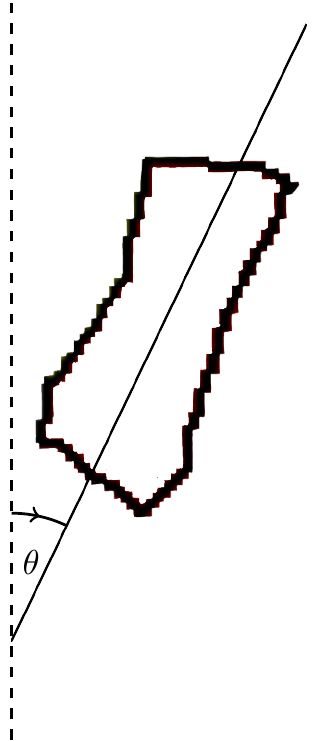}
$\ $
(b)\includegraphics[height=5cm]{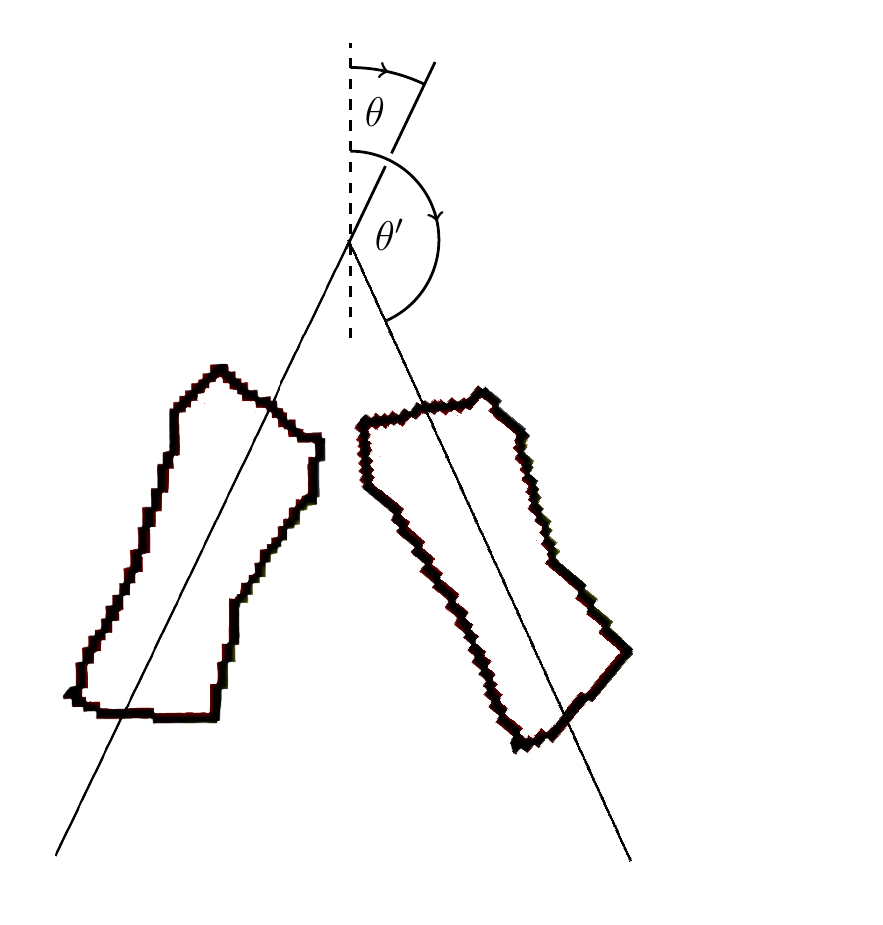}
$\ $
(c)\includegraphics[height=5cm]{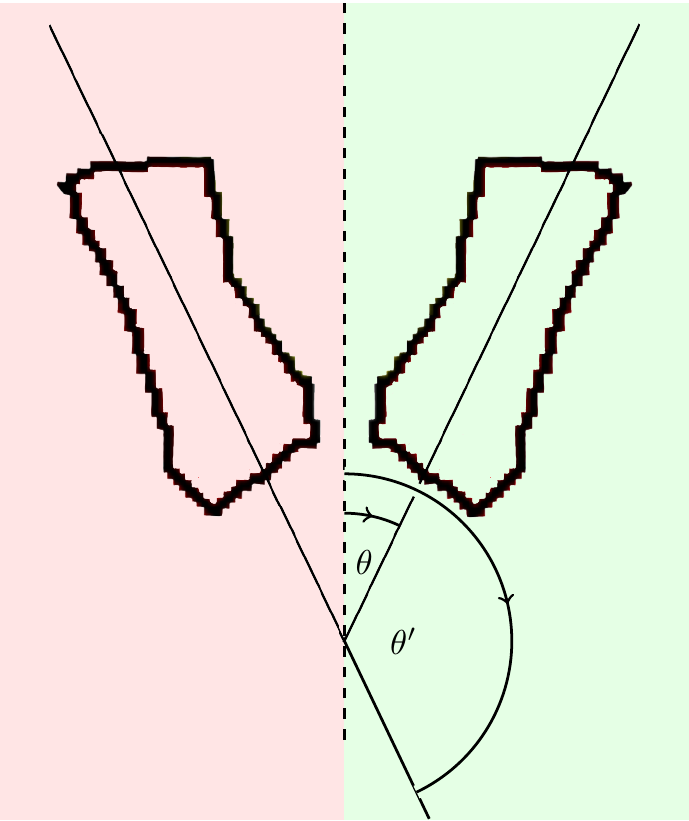}
        \caption{Illustration of the orientation angle $\theta$.
        (a) The orientation angle $\theta\in[0,\pi)$ is measured as the angle between the cell's (major) axis (solid line) and the central axis (dashed line).
        (b) The representation of the orientation ought to be continuous and periodic with period $\pi$, so that an orientation angle of $\theta=\epsilon$ and one of $\theta'=\pi-\epsilon$ continuously map to the same value in the limit $\epsilon\to0$. A large value of $\epsilon$ is shown in the figure to keep the figure legible.
        (c) As far as the morphostate of a cell is concerned, its orientation angle $\theta'=\pi-\theta$ on the left (red) ought be considered to be identical to the orientation angle $\theta$ on the right (green) of the central axis.
        \flabel{orientation}}
\end{figure}

\begin{figure}
\centering
\includegraphics[width=1.0 \textwidth]{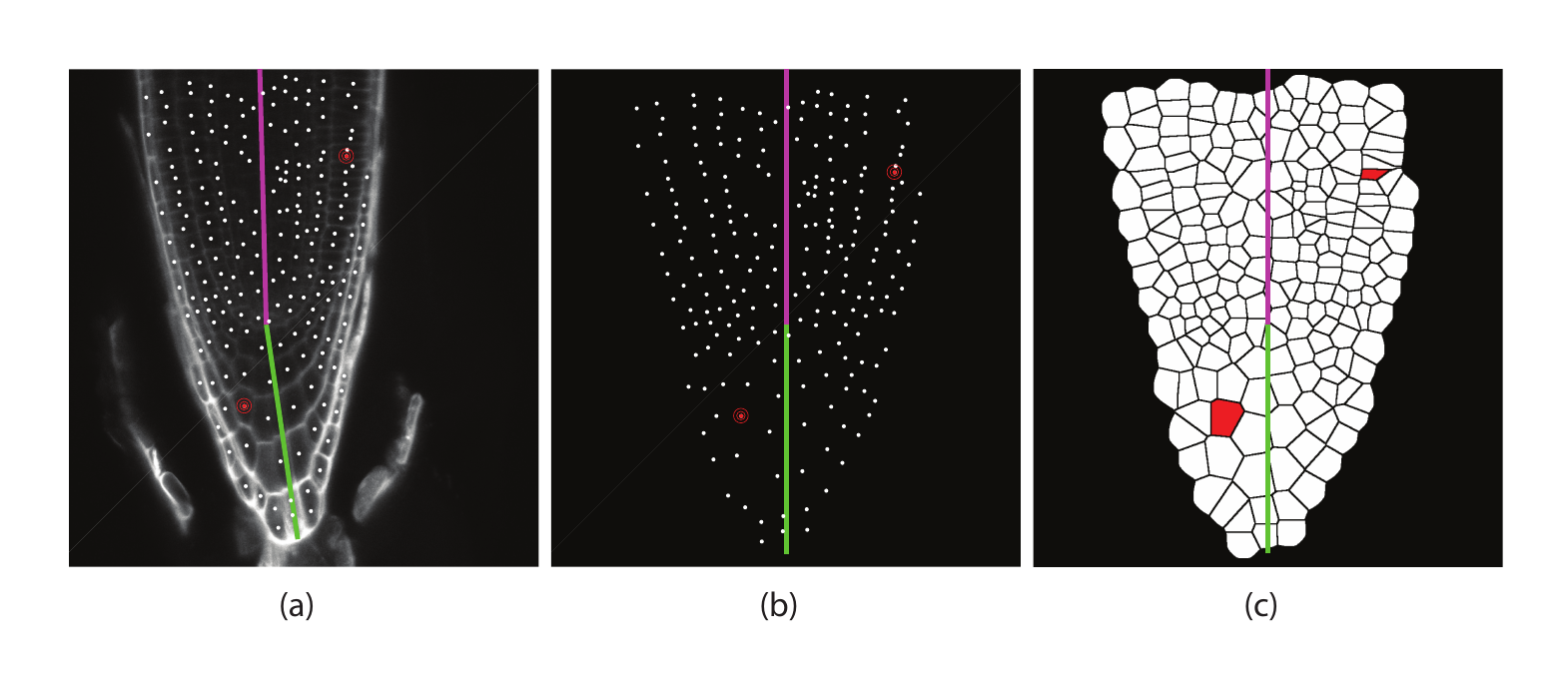}
\caption{\flabel{Registration}
Steps implemented in MATLAB to align (register) roots on top of each other. (a) The root "middle line" is composed of a QC-to-proximal line (purple) and a QC-to-distal one (green); each cell's centroid is rendered with a white dot. Two cells are shown in red to illustrate the registration process. (b) Roto-translation is used to align purple and green portions of the middle line, and centroids are moved to maintain their distance to the middle line. (c) Voronoi tessellation is used to generate virtual cells around centroids.
}
\end{figure}

\end{document}